\documentclass[epj]{svjour}
\usepackage{amsmath,amssymb}
\usepackage{graphicx}% Include figure files
\usepackage{dcolumn}% Align table columns on decimal point
\usepackage{bm}% bold math
\usepackage{bbm}
\usepackage{psfrag}
\usepackage[sort&compress]{natbib}

\newcommand*{\with}{\ensuremath{\quad\text{with}\quad}} 
\newcommand{\I}{\textup{i}}
\newcommand{\E}{\textup{e}}
\newcommand{\D}{\textup{d}}

\newcommand{\pop}[2]{\frac{\partial #1}{\partial #2}}
\newcommand{\poptxt}[2]{\partial #1 / \partial #2}

\newcommand{\ddim}{\udelta\kern0.1em}

\newcommand{\beikonst}[2]{\left( #1 \right)_{\kern-0.2em #2}}
\newcommand{\tr}[2][]{\text{Tr}_{#1}\left\{#2\right\}}
\newcommand{\trtxt}[2][]{\text{Tr}_{#1}\{#2\}}
\newcommand*{\ketL}[1]{\mathopen{|}#1\mathclose{)}}
\newcommand*{\braL}[1]{\mathopen{(}#1\mathclose{|}}
\newcommand*{\sprodL}[2]{\mathopen{(}#1 |#2 \mathclose{)}}

\begin{document}

% -----------------------------------------
%
% Title
%
\title{Heat Conductivity in Small Quantum Systems: Kubo Formula in Liouville Space} %
\author{Mathias Michel\inst{1} \and Jochen Gemmer\inst{2} \and %
        G\"unter Mahler\inst{1}} %
\institute{Institute of Theoretical Physics I, University of Stuttgart, %
             Pfaffenwaldring 57, 70550 Stuttgart, Germany %
\and Physics Department, University of Osnabr\"uck, %
             Barbarastr.\ 7, 49069 Osnabr\"uck, Germany}%
\mail{mathias@theo1.physik.uni-stuttgart.de}%
\date{\today}% 
\abstract{
We consider chains consisting of several identical subsystems weakly coupled by various types of next neighbor interactions.
At both ends the chain is coupled to a respective heat bath with different temperature modeled by a Lindblad formalism.
The temperature gradient introduced by this environment is then treated as an external perturbation.
We propose a method to evaluate the heat current and the local temperature profile of the resulting stationary state as well as the heat conductivity in such systems.
This method is similar to Kubo techniques used e.g.\ for electrical transport but extended here to the Liouville space.
 % end of PACS codes
}

\maketitle

% ---------------------------------------------------------------------------
%
% Main text
%
% ---------------------------------------------------------------------------

% -------------------------------
%
% First Chapter:
%
\section{Introduction}
\label{sec:level1}

As a specific topic of non-equilibrium thermodynamics, heat conduction has been of central interest for a long time.
Instead of reaching a complete equilibrium state, the composite system under some appropriate perturbation enters a local equilibrium state -- small parts of the system approach equilibrium but not the whole system.

Within non-eqilibrium statistical mechanics the theory of linear reponse, originally developed to account for electric conductivity, is a very import method to investigate dynamical as well as static properties of materials \cite{Kubo1991,Mori1956,Mahan1981,Zubarev1974}.
In this context the famous Kubo-formulas \cite{Kubo1957} have led to a rapid developement in the theoretical understanding of processes induced by an external perturbation of the system.
However, a direct mapping of these ideas on pure thermal transport phenomena (perturbations due to thermal gradients \cite{Luttinger1964}) faces serious problems:
Contrary to the case of external perturbations by an electric field, thermal perturbations cannot directly be described by a potential term in the hamiltonian of the system.
Rather, the thermal perturbation is introduced by heat baths with different temperatures coupled to the system, thus calling for a more detailed description than is needed for electric transport.
Nevertheless, those methods are often used, eventually because of their immediate success in describing non-equilibrium processes \cite{Kubo1991,Zotos1997,Heidrich2003,Kluemper2002}.

Recently, the main focus of considerations on heat conduction and Fourier's law has shifted towards small (one dimensional) quantum systems \cite{Saito2003,Saito1996}.
Typically, these systems are chains of identical subsystems weakly coupled by some next neighbor interaction. 
Based on the Lindblad formalism \cite{Lindblad1976} or on techniques of quantum master equations, heat baths are then weakly coupled to the chain at both ends.
The influence of these heat baths enters the Liouville-von-Neumann equation of the system via super operators acting on the density operator of the chain \cite{Saito2002}.
It has been found that in such systems the appearance of a normal heat conduction depends on the type of the interaction between these elementary subsystems \cite{Michel2003}.
Most quantum mechanical interactions show a normal heat conduction behavior (constant non-vanishing local temperature gradients), whereas for special coupling types the local gradient within the chain vanishes (divergence of the conductivity, non-normal scenario).

For a comparison of these two different methods for investigating heat conduction -- the Kubo formulas and the open system approach -- see \cite{Saito2002}.

Instead of solving the full Liouville-von-Neumann equation it should be highly desirable to have a simple but consistent perturbation method available.
However, as already indicated, the mentioned super operators describing the influence of the environment cannot be written as a potential term in the Hamiltonian of the system.
To consistently treat heat currents and local temperature gradients quantitatively also for these model systems, we propose to extend the Kubo technique to Liouville space.

% ---------------------------------------------
%
% Second Chapter:
%
\section{Unperturbed System}
\label{sec:level2}

In the following we will work in the Liouville space of the considered system rather than in the respective Hilbert space.
Thus we consider super operators acting on operators in Hilbert space, e.g.\ the density operator of the system (for a more detailed introduction to super operators see \cite{Schack2000,Tarasov2002,Mukamel2003}).   
The complete Liouville operator of the system under consideration is given here by
\begin{equation}
  \label{eq:1}
  \mathcal{L}= \mathcal{L}_{\text{sys}} 
                + \mathcal{L}_1(T_1) + \mathcal{L}_2(T_2)\;.
\end{equation}
The first term controls the coherent evolution of the quantum system defined by the Hamiltonian $\hat{H}$: 
It is defined by its action on the density operator $\hat{\rho}$ according to
\begin{equation}
  \label{eq:2}
  \mathcal{L}_{\text{sys}} \hat{\rho} 
  = -\frac{\I}{\hbar}\,[\hat{H},\hat{\rho}]\;.
\end{equation}
The system $\hat{H}$ is here a chain of $N$ identical subunits with $n$ levels each, coupled weakly by a next neighbor interaction, thus living in a Liouville space of dimension $n^{2N}$.
One could think of several concrete model systems, for example spin models ($n=2$), for which the Hamiltonian would read
\begin{align}
  \label{eq:20}
  \hat{H} = &\sum_{\mu=1}^{N}\hat{\sigma}_3^{(\mu)}+
  \notag\\
  &\sum_{\mu=1}^{N-1}\big(
  J_x\hat{\sigma}_1^{(\mu)}\hat{\sigma}_1^{(\mu+1)}+
  J_y\hat{\sigma}_2^{(\mu)}\hat{\sigma}_2^{(\mu+1)}+
  J_z\hat{\sigma}_3^{(\mu)}\hat{\sigma}_3^{(\mu+1)}\big)\;.
\end{align}
The first term is the local part of the Hamiltonian, whereas the second defines the interaction between the subsystems ($\hat{\sigma}_i^{(\mu)}$ denote the Pauli operators of the $\mu$th spin).
Choosing $J_x = J_y = J_z$ we get the Heisenberg interaction and for $J_z=0$, $J_x=J_y$ an energy transfer coupling only (XY model).
Furthermore, to avoid any bias we will often use a random next neighbor interaction but without disorder (the same random interaction between different subsystems).

The chain is weakly coupled to two heat baths, one at each end of the system, given by the super operators $\mathcal{L}_1(T_1)$ and $\mathcal{L}_2(T_2)$, where, e.g., in case of spin chains $\mathcal{L}_1(T_1)$ is 
\begin{align}
  \label{eq:30}
  \mathcal{L}_1(T_1)\hat{\rho}
  =& W_1^{\downarrow}(T_1)
     (2 \hat{\sigma}^{(1)}_{-} \hat{\rho} \hat{\sigma}^{(1)}_{+}
     - \hat{\rho} \hat{\sigma}^{(1)}_{+}\hat{\sigma}^{(1)}_{-}
     - \hat{\sigma}^{(1)}_{+}\hat{\sigma}^{(1)}_{-} \hat{\rho})
  \notag\\
  +& W_1^{\uparrow}(T_1)
     (2 \hat{\sigma}^{(1)}_{+} \hat{\rho} \hat{\sigma}^{(1)}_{-}
     - \hat{\rho} \hat{\sigma}^{(1)}_{-}\hat{\sigma}^{(1)}_{+}
     - \hat{\sigma}^{(1)}_{-}\hat{\sigma}^{(1)}_{+} \hat{\rho})\;.
\end{align}
Here, $\hat{\sigma}^{(1)}_{+}$ and $\hat{\sigma}^{(1)}_{-}$ are the raising/lowering operators acting on the first spin, and the $W$'s are the so called rates which contain the bath temperature.
These are standard Lindblad operators \cite{Lindblad1976}, well known from the theory of open systems in quantum optics, which introduce the damping of the environement into the Liouville-von-Neumann equation of the system.
This procedure is the same as a quantum master equation \cite{Nakajima1958,Zwanzig1960}, used to describe dissipative quantum systems, for example in \cite{Miyashita2003}.
They can easily be generalised to finite $n>2$.
$\mathcal{L}_2(T_2)$ is defined correspondingly.

For the unperturbed system both bath systems are at the same temperature $T_1=T_2=T$.
The whole Liouville-von-Neumann equation of the unperturbed system then reads
\begin{equation}
  \label{eq:3}
  \pop{}{t}\, \hat{\rho} = \mathcal{L}_0 \hat{\rho}\;,
\end{equation}
with
\begin{equation}
  \label{eq:28}
  \mathcal{L}_0 = \mathcal{L}_{\text{sys}}
                + \mathcal{L}_1(T) + \mathcal{L}_2(T)\;.
\end{equation}
Since the two baths have exactly the same temperature we expect the system to settle in a thermal stationary state $\hat{\rho}_0$:
This state should support neither a heat current nor temperature gradients -- it is a global equilibrium state with temperature $T$.

The eigenvalues and eigenvectors of the unperturbed system are given by the eigen equation
\begin{equation}
  \label{eq:4}
  \mathcal{L}_0 \ketL{\hat{\rho}_j} = l_j \ketL{\hat{\rho}_j}
  \;,\;j=0,\dots,n^{2N}-1\;,
\end{equation}
where the ket-vectors in Liouville space have been denoted as $\ketL{\dots}$.
A scalar product of vectors in Liouville space can be defined by
\begin{equation}
  \label{eq:25}
  \sprodL{\hat{\rho}_i}{\hat{\rho}_j}
  = \tr{\hat{\rho}_i^{\dagger}\hat{\rho}_j}\;.
\end{equation}
The (unique) stationary state $\hat{\rho}_0$ is also an eigenvector of the system with eigenvalue zero, $\mathcal{L}_0 \ketL{\hat{\rho}_0}=0$, whereas all other eigenvalues have a negative real part.
This is due to the fact that asymptotically the system should enter the equilibrium state $\ketL{\hat{\rho}_0}$ regardless of which state the system was at the beginning.
No other eigenvector is able to contribute to the equilibrium state, i.e.\ all other eigenvectors must be unstable.

Since the Liouville operator $\mathcal{L}_0$ is not hermitian, the eigenvectors do not form an orthogonal basis, i.e.\
\begin{equation}
  \label{eq:5}
  \sum_j \ketL{\hat{\rho}_j}\braL{\hat{\rho}_j}=\mathcal{G}
\end{equation}
is in general not the unit operator in Liouville space (c.f.\ \cite{Schack2000}).
But with the help of the super operator $\mathcal{G}$ it is possible to find a dual basis $\ketL{\hat{\rho}^j} = \mathcal{G}^{-1}\ketL{\hat{\rho}_j}$ with the property
\begin{equation}
  \label{eq:6}
  \sum_j \ketL{\hat{\rho}_j}\braL{\hat{\rho}^j} = \mathbbm{1}\;.
\end{equation}
Based on this complete basis, we can expand each state of the system in terms of the eigensystem 
\begin{equation}
  \label{eq:35}
  \ketL{\hat{\rho}} 
  = \sum_{j=0}^{n^{2N}-1} \ketL{\hat{\rho}_j}\sprodL{\hat{\rho}^j}{\hat{\rho}}
  = \ketL{\hat{\rho}_0} + \sum_{j=1}^{n^{2N}-1} c_j \ketL{\hat{\rho}_j}\;,
\end{equation}
with $c_j = \sprodL{\hat{\rho}^j}{\hat{\rho}}$.
Because the system will asymptotically be found in the normalized stationary state $\ketL{\hat{\rho}_0}$ and we have to require that the state $\ketL{\hat{\rho}}$ is normalized for the whole time evolution, each eigenstate must be tracefree (except $\ketL{\hat{\rho}_0}$).

% ---------------------------------------
%
% Third Chapter:
%
\section{Perturbation and Local Equilibrium State}
\label{sec:level3}

The system will be perturbed now by applying a small temperature gradient $\Delta T$. 
We start at time $t=-\infty$ and switch on the perturbation exponentially till $t=0$, letting $\Delta T$ constant for all times $t>0$.
Thereafter the system is subject to this small constant external temperature gradient.
We are interested in the properties of the stationary \emph{local} equilibrium state of the system reached in the limit $t\rightarrow\infty$ (this state now contains stationary currents and a constant temperature profile).

The Liouville operator of the perturbation is thus given by
\begin{equation}
  \label{eq:7}
  \mathcal{L'}(\Delta T, t) = \mathcal{L}_1(T+\frac{\Delta T}{2}f(t)) 
                            + \mathcal{L}_2(T-\frac{\Delta T}{2}f(t))\;,
\end{equation}
where
\begin{equation}
  \label{eq:8}
  f(t) = \Theta(-t)\E^{t}+\Theta(t)\;, \with
  \Theta(t)=
  \begin{cases}
  0 & t < 0 \\
  1 & t > 0
  \end{cases}\;.
\end{equation}
The two environment operators are the same as before but with a time dependent external temperature difference.

The time evolution of the whole system under the influence of the perturbation is given by the Liouville-von-Neumann equation
\begin{equation}
  \label{eq:9}
  \pop{}{t} \hat{\rho} = (\mathcal{L}_0 + \mathcal{L'}(\Delta T, t))\hat{\rho}
  \;.
\end{equation}
Starting from the past in a thermal equilibrium state $\hat{\rho}_0$, we assume the time dependent state of the whole system to be 
\begin{equation}
  \label{eq:10}
  \hat{\rho} = \hat{\rho}_0 + \Delta\hat{\rho}(t)\;.
\end{equation}
Introducing this into (\ref{eq:9}), supressing terms of higher oder in the perturbation and observing that $\poptxt{\hat{\rho}_0}{t}=\mathcal{L}_0\hat{\rho}_0=0$ one finds the time evolution equation for $\Delta\hat{\rho}(t)$,
\begin{equation}
  \label{eq:11}
  \pop{}{t}\Delta\hat{\rho}(t) - \mathcal{L}_0 \Delta\hat{\rho}(t)
  = \mathcal{L'}(\Delta T, t) \hat{\rho}_0\;.
\end{equation}
With a transformation similar to the one introduced by Kubo for the Hilbert-space
\begin{equation}
  \label{eq:12}
  \E^{\mathcal{L}_0 t} 
  \Big(\pop{}{t}\big(\E^{-\mathcal{L}_0 t}\Delta\hat{\rho}(t)\big) \Big)
  = \pop{}{t}\Delta\hat{\rho}(t) - \mathcal{L}_0 \Delta\hat{\rho}(t)\;,
\end{equation}
one can integrate the differential equation, finding the formal solution for the time-dependent perturbation
\begin{equation}
  \label{eq:13}
  \Delta\hat{\rho}(t) 
  = \int_{-\infty}^{t} \D t' \E^{\mathcal{L}_0 (t-t')}
    \mathcal{L'}(\Delta T, t') \hat{\rho}_0\;.
\end{equation}
Let us call this equation the Kubo-formula in Liouville space (see also \cite{Kubo1991}).

In the case of a chain of two level systems each of the two super operators $i=1,2$ of the bath coupling at both ends of the system consist of two transition processes (in case of finite temperatures)
\begin{equation}
  \label{eq:14}
  \mathcal{L}_i(T) = W_i^{\downarrow}(T) \mathcal{E}_i^{\downarrow}
                   + W_i^{\uparrow}(T) \mathcal{E}_i^{\uparrow}\;,
\end{equation}
with the two rates $W_i^{\downarrow}(T)=(1-T)\lambda_{\text{B}}$ and $W_i^{\uparrow}(T)=T\lambda_{\text{B}}$ ($\lambda_{\text{B}}$ is the coupling strength of the environment, $T$ its temperature) and $\mathcal{E}_{i}^{\downarrow}$, $\mathcal{E}_{i}^{\uparrow}$ are transition operators.
According to (\ref{eq:30}) these transition operators read, e.g.,
\begin{equation}
  \label{eq:29}
  \mathcal{E}_1^{\downarrow} \hat{\rho} 
  =  (2 \hat{\sigma}^{(\mu)}_{-} \hat{\rho} \hat{\sigma}^{(\mu)}_{+}
     - \hat{\rho} \hat{\sigma}^{(\mu)}_{+}\hat{\sigma}^{(\mu)}_{-}
     - \hat{\sigma}^{(\mu)}_{+}\hat{\sigma}^{(\mu)}_{-} \hat{\rho})
\end{equation}
(For more energy levels of the subunits in the chain one has to account for all possible transitions -- a straight-forward extension.)

With this definition the perturbed super operator (\ref{eq:7}) can be rewritten as
\begin{equation}
  \label{eq:15}
  \mathcal{L'}(\Delta T,t) 
  = \mathcal{L}_{1}(T)+\mathcal{L}_{2}(T)
  + \frac{\Delta T \lambda_{\text{B}}}{2} f(t) \mathcal{E}\;,
\end{equation}
with $\mathcal{E}=-\mathcal{E}_{1}^{\downarrow}+\mathcal{E}_{1}^{\uparrow}+\mathcal{E}_{2}^{\downarrow}-\mathcal{E}_{2}^{\uparrow}$.
The first two terms just replicate the bath operators with the same temperature at both ends. 
Acting on the unperturbed equilibrium state these terms vanish and therefore (\ref{eq:13}) reduces to
\begin{equation}
  \label{eq:16}
  \Delta\hat{\rho}(t) 
  = \frac{\Delta T\lambda_{\text{B}}}{2}
    \int_{-\infty}^{t} \D t' \E^{\mathcal{L}_0 (t-t')}
    f(t') \mathcal{E} \hat{\rho}_0\;.
\end{equation}
Introducing the unit operator of the Liouville space, defined in (\ref{eq:6}), into (\ref{eq:16})
\begin{equation}
  \label{eq:17}
  \Delta\hat{\rho}(t) 
  = \frac{\Delta T\lambda_{\text{B}}}{2}
    \int_{-\infty}^{t} \D t' \E^{\mathcal{L}_0 (t-t')}
    \sum_j \ketL{\hat{\rho}_j}\braL{\hat{\rho}^j}
    f(t') \mathcal{E} \hat{\rho}_0\;,
\end{equation}
we find using $\E^{\mathcal{L}_0 (t-t')}\ketL{\hat{\rho}_j}=\E^{l_j(t-t')}\ketL{\hat{\rho}_j}$
\begin{equation}
  \label{eq:18}
  \Delta\hat{\rho}(t) 
  = \frac{\Delta T\lambda_{\text{B}}}{2}
    \sum_j \braL{\hat{\rho}^j} \mathcal{E} \ketL{\hat{\rho}_0}
    \ketL{\hat{\rho}_j}
    \int_{-\infty}^{t} \E^{l_j(t-t')} f(t') \D t'\;.
\end{equation}
By integrating over $t'$ with the function $f(t')$ as defind in (\ref{eq:8}) and taking into account that the real part of $l_j$ is negative (for $j\neq 0$), we finally get the time dependent perturbation of the density operator
\begin{align}
  \label{eq:19}
  \Delta\hat{\rho}(t) 
  =& \frac{\Delta T\lambda_{\text{B}}}{2}
    \Big( (1+t) \braL{\hat{\rho}^0} \mathcal{E} \ketL{\hat{\rho}_0}
          \ketL{\hat{\rho}_0}
  \notag\\
         &+\sum_{j=1}^{n^{2N}-1}
          \big(\frac{\E^{l_j t}}{1-l_j}+\frac{\E^{l_j t}-1}{l_j}\big)
          \braL{\hat{\rho}^j} \mathcal{E} \ketL{\hat{\rho}_0}
          \ketL{\hat{\rho}_j}
    \Big)\;.
\end{align}
The first term results from the integration over the addend $j=0$, the sum contains the rest. 
Rewriting the matrix element of the super operator $\mathcal{E}$ as a scalar product defined by (\ref{eq:25}), we find
\begin{equation}
  \label{eq:24}
  \braL{\hat{\rho}^0} \mathcal{E} \ketL{\hat{\rho}_0} 
  =  \sprodL{\hat{\rho}^0}{\mathcal{E}\hat{\rho}_0}
  = \tr{\hat{\rho}_0 (\mathcal{E} \hat{\rho}_0)}\;.
\end{equation}
The conservation of the trace of the complete Liouville-von-Neumann equation (\ref{eq:9}) requires that each operator produced by an action of the perturbation operator $\mathcal{E}$ on an arbitrary density operator has to be a trace free operator.
Since $\ketL{\hat{\rho}_0}$ is the only basis state with non-zero trace, as argued before, we have to require that $\mathcal{E} \hat{\rho}_0$ has no component in $\ketL{\hat{\rho}_0}$ direction.
Therefore this matrix element should be zero and (\ref{eq:19}) reduces to
\begin{equation}
  \label{eq:26}
  \Delta\hat{\rho}(t) 
  =\frac{\Delta T\lambda_{\text{B}}}{2}
   \sum_{j=1}^{n^{2N}-1}
   \big(\frac{\E^{l_j t}}{1-l_j}+\frac{\E^{l_j t}-1}{l_j}\big)
   \braL{\hat{\rho}^j} \mathcal{E} \ketL{\hat{\rho}_0}
   \ketL{\hat{\rho}_j}\;.
\end{equation}
This perturbative term includes all currents and local temperature gradients of the system under the perturbation. 
Since we are interested in a local equilibrium state -- a stationary state with a constant current and temperature profile, which will be reached after a certain relaxation time, we consider (\ref{eq:26}) in the limit of $t\rightarrow\infty$ finding
\begin{equation}
  \label{eq:27}
  \Delta\hat{\rho} = \lim_{t\rightarrow\infty} \Delta\hat{\rho}(t)
  = -\frac{\Delta T\lambda_{\text{B}}}{2}
    \sum_{j=1}^{n^{2N}-1}
    \frac{\braL{\hat{\rho}^j} \mathcal{E} \ketL{\hat{\rho}_0}}{l_j}\,
    \ketL{\hat{\rho}_j}\;.
\end{equation}
This is the first-order change of the density operator introduced by the perturbation.

% ---------------------------------------
%
% Forth Chapter:
%
\section{Current and Local Temperature Profile}
\label{sec:level4}

Now we are able to account for the local temperature profile and the expectation value of the current.
The energy current operator $\hat{J}^{(\mu,\mu+1)}$ can be derived from a discretized version of the continuity equation (formulating a current into and out of the site $\mu$, respectively)
\begin{equation}
  \label{eq:32}
  \I [ \hat{H}, \hat{H}^{(\mu)}_{\text{loc}} ]
  = \hat{J}^{(\mu-1,\mu)} - \hat{J}^{(\mu,\mu+1)}\;.
\end{equation}
As a measure for the temperature $T$ of a single subsystem we use here the local energy of the system so that $0\leq T \leq 0.5$ in units of the local level spacing, c.f.\ \cite{Michel2003,Gemmer2004}.
This should be appropriate for weakly coupled subsystems within the chain (only a very small amount of energy is within the interaction).
The operator 
\begin{equation}
  \label{eq:31}
  \Delta\hat{H}^{(\mu,\mu+1)}_{\text{loc}} 
  = \hat{H}^{(\mu)}_{\text{loc}} - \hat{H}^{(\mu+1)}_{\text{loc}}\;,
\end{equation}
measures the local energy difference between two adjacent subsystems $\mu$ and $\mu+1$ ($\mu=1,2,\dots,N-1$). 

Since the whole stationary density operator of the system is now given by $\hat{\rho}=\hat{\rho}_0+\Delta\hat{\rho}$ and since we know that $\hat{\rho}_0$ does not give rise to any local temperature difference and current, the expectation values of the operators defined above are determined only by $\Delta\hat{\rho}$ from (\ref{eq:27}).
Therefore we find for the local internal temperature gradient 
\begin{align}
  \label{eq:23}
  &\delta T^{(\mu,\mu+1)} 
   = \trtxt{\Delta\hat{H}^{(\mu,\mu+1)}_{\text{loc}} \Delta\hat{\rho}}\notag\\
  &\quad= -\frac{\Delta T\lambda_{\text{B}}}{2} \sum_{j=1}^{n^{2N}-1}
     \frac{\braL{\hat{\rho}^j} \mathcal{E} \ketL{\hat{\rho}_0}}{l_j}\,
     \trtxt{\Delta\hat{H}^{(\mu,\mu+1)}_{\text{loc}}\hat{\rho}_j}
\end{align}
and the local current within the system
\begin{align}
  \label{eq:33}
  &J^{(\mu,\mu+1)} 
   = \tr{\hat{J}^{(\mu,\mu+1)} \Delta\hat{\rho}}\notag\\
  &\quad= -\frac{\Delta T\lambda_{\text{B}}}{2} \sum_{j=1}^{n^{2N}-1}
     \frac{\braL{\hat{\rho}^j} \mathcal{E} \ketL{\hat{\rho}_0}}{l_j}\,
     \trtxt{\hat{J}^{(\mu,\mu+1)}\hat{\rho}_j}\;.
\end{align}
The current as well as the local temperature gradient are thus found to depend linearly on the global temperature difference of the bath systems.
Under stationary conditions the current must be independent of $\mu$, $J^{(\mu,\mu+1)}=J$, so that (\ref{eq:33}) can be rewritten as
\begin{equation}
  \label{eq:34}
  J = - \kappa' \Delta T\;.
\end{equation}
Eigenstates and eigenvalues entering here the \emph{global} conductivity $\kappa'$ depend only on the mean temperature of the unperturbed system, not on $\Delta T$.
Based on this $\kappa'$ as a global property of the system, including its contact properties to the environments, let us call (\ref{eq:34}) ``external Fourier's Law''.  

Furthermore, combining (\ref{eq:23}) and (\ref{eq:33}), we can define a \emph{local} conductivity within the system 
\begin{equation}
  \label{eq:21}
  \kappa^{(\mu,\mu+1)} 
  = -\frac{J^{(\mu,\mu+1)}}{\delta T^{(\mu,\mu+1)}}
  = -\frac{J}{\delta T^{(\mu,\mu+1)}}
\end{equation}
implying also $\kappa^{(\mu,\mu+1)}$ to be independent of the external gradient $\Delta T$.

\begin{figure}
  \centering  
  \psfrag{kappa}{$\kappa^{(2,3)}$}
  \psfrag{DeltaT}{\raisebox{-4pt}{$\Delta T$}}
  \psfrag{das ist der erste}{\hspace{-2mm}\tiny eq. (\ref{eq:9})}
  \psfrag{das ist der zweite}{\hspace{-2mm}\tiny eq.\ (\ref{eq:21})}
  \includegraphics[width=5cm]{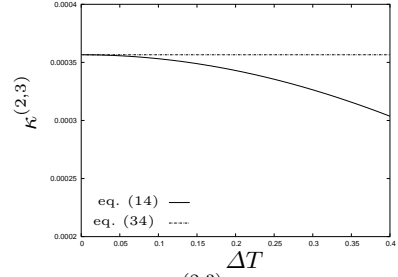}
  \caption{Local conductivity $\kappa^{(2,3)}$ in a Heisenberg spin chain of 4 spins as a function of the external perturbation $\Delta T$; the solid line refers to the solution of the full Liouville-von-Neumann equation, the dashed line shows (\ref{eq:21}).}
\label{fig:1}  
\end{figure}

We can compare this result with the complete numerical solution of the Liouville-von-Neumann equation (\ref{eq:9}), here for a Heisenberg spin chain with four spins.
In Fig.~\ref{fig:1} we show the local conductivity of the two central spin systems $\kappa^{(2,3)}$ as a function of the external gradient $\Delta T$.
Indeed, we find numerically that $\kappa^{(2,3)}$ according to (\ref{eq:21}) does not depend on $\Delta T$ (dashed line).
The exact numerical solution of the Liouville-von-Neumann equation shows a weak $\Delta T$ dependence (solid line).
As expected, linear transport and our perturbational theory applies for not too large external gradients $\Delta T$ only.
Note that the full range of temperatures $0\leq T<\infty$ has been mapped here onto the interval $[0,0.5]$ ($\Delta T = 0.3$ is thus already a large gradient). 

The theoretical predictions of the perturbation theory at hand concerning temperature gradients and currents within the chain for all investigated systems (Heisenberg, XY and random coupling model) are in very good accordance with the numerical solution of the complete Liouville-von-Neumann equation of the system (compare \cite{Michel2003}).

``Normal'' heat conduction (Fourier's Law) is associated with a constant but non-zero local temperature gradient and thus a finite conductivity $\kappa^{(\mu,\mu+1)}=\kappa$ independent of site $\mu$ everywhere in the system (see \cite{Mahan1981,Gemmer2004}).
But note that this internal conductivity $\kappa$ is not necessarily identical with the \emph{global} $\kappa'$ since the latter depends additionally on the bath contacts. 
In the limit of long chains the conductivity should be independent of the contacts, yielding the correct scaling behavior within the chain (see \cite{Michel2003}).

The majority of coupling types within the chain, the Heisenberg coupling and the random next neighbor interaction, indeed show this normal behavior in the weak coupling limit \cite{Michel2003,Gemmer2004}.
But this ``normal'' transport type does not always show up:
A spin-spin interaction consisting of an energy transfer coupling only (XY model) leads to a vanishing temperature gradient off the contact regions.
This vanishing gradient implies a divergent conductivity within the chain $\kappa^{(\mu,\mu+1)}$.
Nevertheless the current remains finite because of the resistance at the contacts, therefore the global conductivity $\kappa$ defined in (\ref{eq:34}) also remains finite for this special coupling type.
Therefore we could state that the ``external Fourier's Law'' is valid even if Fourier's Law proper does not apply. 
These results can be compared with the numerical results of the full solution of the Liouville-von-Neumann equation and we find perfect agreement.

% ----------------------------------------------------------------------------
%
% 5. Chapter:
%
\section{Conclusion}
\label{sec:level5}

We have considered heat conduction in small quantum systems built up from identical subsystems weakly coupled by some next neighbor interaction.
By a perturbation theory similar to that introduced by Kubo but extended to the full Liouville space of the system, we have been able to derive a quantitative equation for the temperature gradient and heat currents within such systems. 
These equations  depend only on properties of the unperturbed system and linearly on the strength of the perturbation $\Delta T$.

The most remarkable point of the equation for the heat current and the temperature profile is the fact that the global temperature difference of the external bath systems shows up only as a parameter.
This is not only a numerical advantage, since a diagonalization for different global gradients is no longer necessary, but also an interesting physical fact:
The heat conductivity is independent of the external gradient $\Delta T$, and the ``external Fourier's Law'' is always fulfilled, even if the internal gradient of the system is not constant, as long as the perturbation theory applies. 

Our approach does not have the problem of introducing a potential term into the Hamiltonian of the system, like in standard Kubo formulas for heat conduction.
The bath systems, modeled by a Linblad formalism, directly define the perturbation in Liouville space.
Like in standard perturbation theory in Hilbert space, the first order correction to the stationary state of the system is expressed in terms of transition matrix elements of the perturbation operator and the eigenstates and eigenvalues of the unperturbed system.
Only the non-orthogonality of the eigensystem of the unperturbed system needs a more careful treatment, formally the equations are very similar. 

In the future we intend to investigate further aspects of the derived formalism, hoping to clarify the question of the different transport behavior (non-vanishing and vanishing local gradients) under different coupling types.

% -----------------------------------------------------------------------------
%
% The End of the Main Text
%
% -----------------------------------------------------------------------------

\begin{acknowledgement}
We thank M.\ Hartmann, M.\ Henrich, Ch.\ Kostoglou, H. Michel, H.\ Schmidt, M.\ Stollsteimer and F. Tonner for fruitful discussions. Financial support by the Deutsche Forschungsgesellschaft is gratefully acknowledged.
\end{acknowledgement}


\begin{thebibliography}{10}

\bibitem{Kubo1991}
R.~Kubo, M.~Toda, and N.~Hashitsume.
\newblock {\em Statistical {P}hysics {II}: {N}onequilibrium {S}tatistical
  {M}echanics}.
\newblock Number~31 in Solid-State Sciences. Springer, Berlin, Heidelberg,
  New-York, 2. edition, 1991.

\bibitem{Mori1956}
H.~Mori.
\newblock Correlation {F}unction {M}ethod for {T}ransport {P}henomena.
\newblock {\em Phys. Rev.}, 115(2):298--300, 1956.

\bibitem{Mahan1981}
G.~D. Mahan.
\newblock {\em Many-{P}article {P}hysics}.
\newblock Plenum Press, New York, London, 3. edition, 2000.

\bibitem{Zubarev1974}
D.~N. Zubarev.
\newblock {\em Nonequilibrium {S}tatistical {T}hermodynamics}.
\newblock Studies in Soviet Sciences. Consultants Bureau, New York, London,
  1974.
\newblock transl. by P. J. Shepherd.

\bibitem{Kubo1957}
R.~Kubo.
\newblock Statistical-{M}echanical {T}heory of {I}rreversible {P}rocesses. {I}.
\newblock {\em J. Phys. Soc. Jpn.}, 12:570--586, 1957.

\bibitem{Luttinger1964}
J.~M. Luttinger.
\newblock Theory of {T}hermal {T}ransport {C}oefficients.
\newblock {\em Phys. Rev.}, 135(6A):A1505--A1514, 1964.

\bibitem{Zotos1997}
X.~Zotos, F.~Naef, and P.~Prelovsek.
\newblock Transport and conservation laws.
\newblock {\em Phys. Rev. B}, 55(17):11029--11032, 1997.

\bibitem{Heidrich2003}
F.~Heidrich-Meisner, A.~Honecker, D.C. Cabra, and W.~Brenig.
\newblock Zero-{F}requency transport properties of one dimensional spin-1/2
  systems.
\newblock {\em cond-math/0304595}, 2003.

\bibitem{Kluemper2002}
A.~Kl\"umper and K.~Sakai.
\newblock The thermal conductivity of the spin-1/2 {XXZ} chain at arbitrary
  temperature.
\newblock {\em J. Phys. A: Math. Gen.}, 35:2173--2182, 2002.

\bibitem{Saito2003}
K.~Saito.
\newblock Strong evidence of normal heat conduction in a one-dimensional
  quantum system.
\newblock {\em Europhys. Lett.}, 61:34--40, 2003.

\bibitem{Saito1996}
K.~Saito, S.~Takesue, and S.~Miyashita.
\newblock Thermal conduction in a quantum system.
\newblock {\em Phys. Rev. E}, 54:2404--2408, 1996.

\bibitem{Lindblad1976}
G.~Lindblad.
\newblock On the {G}enerators of {Q}uantum {D}ynamical {S}emigroups.
\newblock {\em Commun. Math. Phys.}, 48:119--130, 1976.

\bibitem{Saito2002}
K.~Saito and S.~Miyashita.
\newblock Enhancement of thermal conductivity in gapped quantum spin chains.
\newblock {\em J. Phys. Soc. Jpn.}, 71:2485--2488, 2002.

\bibitem{Michel2003}
M.~Michel, M.~Hartmann, J.~Gemmer, and G.~Mahler.
\newblock Fourier's {L}aw confirmed for a class of small quantum systems.
\newblock {\em Euro. Phys. J. B}, 34:325--330, 2003.

\bibitem{Schack2000}
R.~Schack and M.~C. Caves.
\newblock Explicit product ensembles for separable quantum states.
\newblock {\em J. Mod. Opt.}, 47(2/3):387--399, 2000.

\bibitem{Tarasov2002}
V.E. Tarasov.
\newblock Pure states of open quantum systems.
\newblock {\em Phys. Rev. E}, 66:056116, 2002.

\bibitem{Mukamel2003}
S.~Mukamel.
\newblock Superoperator representation of nonlinear response: {U}nifying
  quantum field and mode coupling theories.
\newblock {\em Phys. Rev. E}, 68(2):021111, 2003.

\bibitem{Nakajima1958}
S.~Nakajima.
\newblock On quantum theory of transport phenomena.
\newblock {\em Prog. Theor. Phys.}, 20:948--959, 1958.

\bibitem{Zwanzig1960}
R.~Zwanzig.
\newblock Ensemble {M}ethod in the {T}heory of {I}rreversibility.
\newblock {\em J. Chem. Phys.}, 33:1338, 1960.

\bibitem{Miyashita2003}
S.~Miyashita and K.~Saito.
\newblock Real time scale simulation for quantum processes in dissipative
  environments.
\newblock {\em Physica B}, 329-333:1142--1143, 2003.

\bibitem{Gemmer2004}
J.~Gemmer, M.~Michel, and G.~Mahler.
\newblock {\em Quantum {T}hermodynamics}, volume 657 of {\em LNP}.
\newblock Springer, Heidelberg, Berlin, 2004.
\newblock in press.

\end{thebibliography}
\end{document}